\documentclass[%
 reprint,
 amsmath,amssymb,
 aps,
]{revtex4-1}

\usepackage{graphicx}
\usepackage{dcolumn}
\usepackage{bm}
\usepackage{color}%

\begin{document}

\preprint{APS/123-QED}

\title{Super-intense Single Attosecond Pulse Generation by Plasma Gating}

\author{Suo Tang}
\email{suo.tang@mpi-hd.mpg.de}
\author{Naveen Kumar}
\affiliation{Max~Planck~Institute for Nuclear Physics, Saupfercheckweg 1, 69117 Heidelberg, Germany}%

\date{\today}

\begin{abstract}
A robust plasma gating to generate a single ultra-intense attosecond pulse is developed. It is a manifestation of the hole-boring effect that limits the strongest attosecond pulse emission within one laser cycle. The generated pulse is characterized by a stabilized harmonic phase $\psi \approx \pm\pi/2$ and a slowly decaying exponential spectrum bounded by $\gamma$-spike scaling and CSE scaling. The phase oscillations in low-frequency region and fluctuations in high-frequency region are discussed. We also show that the phase fluctuations in high-frequency region can be reduced by including radiation reaction force.

\end{abstract}

\pacs{Valid PACS appear here}
\maketitle


The field of attosecond (AS) metrology is an emerging area of research spanning a range of applications from atomic physics to biological sciences~\cite{Krausz:2014aa}. An AS pulse can be regarded as a hyperfast camera that can freeze the motion of electrons, thus making it an invaluable tool to study many fundamental physical processes in real-time (AS spectroscopy)~\cite{Krausz:2009aa}. Due to the difficulties involved in generating an ultra-intense single AS pulse, the application of the pumb-probe AS spectroscopy is currently limited~\cite{Krausz:2009aa,*Kienberger:2002aa}. Hence the generation of an ultra-intense AS pulse with stable carrier-envelope phase (CEP) can open the hitherto unexplored regime of nonlinear AS spectroscopy~\cite{Ledingham:2003aa,*Milosevic:2004aa}, extending the AS metrology to high-energy quantum electrodynamical processes~\cite{Di-Piazza:2012aa,*Marklund:2006aa}.

To generate an AS pulse, one has to significantly broaden, via nonlinear processes, the frequency spectrum of a femtosecond laser pulse. This frequency broadening is termed as high-order harmonic generation (HHG) and can be accomplished via the interaction of a strong laser pulse with either gaseous or solid targets. The underlying physical mechanisms for both cases have been extensively studied. Though the generation of a single AS pulse has been experimentally demonstrated via the interaction of a laser with a gas jet target, the efficiency saturates at relativistic laser intensities~\cite{Teubner:2009aa}. On the other hand, HHG from the solid target scales favorably at relativistic intensities and has been studied both theoretically and experimentally, \emph{e.g.}  relativistically oscillating mirror (ROM), coherent wake emission (CWE), coherent synchrotron emission (CSE) and relativistic electron spring (RES)~\cite{Bulanov:2013aa,Lichters:1996ab,Tarasevitch:2000aa,Baeva:2006aa,Brugge:2010aa,Thaury:2010aa,Gonoskov:2011aa,PRL245005CSE,Dromey:2006aa,Dromey:2012aa,Nomura:2009aa,Heissler:2012aa,Rodel:2012aa,Behmke:2011aa}. The solid HHG usually results (after filtering a range of frequency components) in a train of AS pulses~\cite{Tsakiris:2006aa}. To generate an isolated AS pulse, several techniques, \emph{e.g.} polarization and intensity gatings~\cite{Tsakiris:2006aa}, attosecond lighthouse effect~\cite{Vincenti:2012aa} and focusing of the harmonic from $\lambda^3$ ($\lambda$ is the wavelength of the laser pulse) focal-spot size volume~\cite{Naumova:2004aa} have been implemented. However, the implementation of these techniques to ultra-relativistic laser intensities, $a_0=eE_{l}/m_{e}c\omega_{l} \gg 1$, may not be practical due to severe constraints on laser pulse and target parameters \emph{e.g.} stable CEP, few cycle laser pulses, extremely thin ($l\ll \lambda_l$) targets etc, where $e$ and $m_{e}$ denote the electron charge and rest mass, $E_{l}$ and $\omega_{l}$ are the laser electric field and frequency, $c$ is the light speed in vacuum. Moreover, additional physical processes$-$which had not been properly addressed in previous literatures$-$such as hole-boring (HB) and radiation reaction (RR) effects~\cite{Landau:2005aa} also become important in the ultra-relativistic laser-solid interaction. Recently, we showed that HB effect can broaden the harmonic peaks leading to significant overlap between harmonics in the frequency domain~\cite{Tang:2017aa}. This harmonic overlap is the prerequisite for the generation of a single AS pulse. We also showed that radiation reaction force affects the intensity of the harmonics~\cite{Tang:2017aa}.

In this Letter, we develop a new mechanism for the generation of an isolated super-intense AS pulse. We show that the HB effect can effectively limit the strongest pulse emission within one laser cycle, making it possible to isolate an AS pulse. We term this HB induced pulse isolation as a \textquotedblleft Plasma Gate \textquotedblright. This mechanism is indeed dominant in the ultra-relativistic regime and works for long laser pulses. The harmonics constituting the generated pulse are phase-locked to $\psi \approx \pm\pi/2$ due to the dynamics of the plasma surface electron layer. For generality, we consider a fully ionized plasma with a pre-plasma, $n_{e}=n_c/2\exp{(x/L)}$, where $n_{c}=\omega_L^2m_e/4\pi e^2$ is the plasma critical density. The incidence of a linearly polarized laser pulse onto a solid target is simulated in $1$D geometry with the EPOCH-PIC code~\cite{Arber:2015aa}. RR force is included by employing Landau-Lifshitz prescription~\cite{Landau:2005aa}. Plasma collisions are included in all cases. From now onwards, we use the dimensionless quantities: $n_e=n_e/n_{c}$, $t=\omega_{l}t$, $x=k_{l}x$, $\beta=v/c$, $\omega=\omega/\omega_{l}$, $E=eE/(m_{e}c\omega_{l})$.

\begin{figure}
  \includegraphics[width=0.45\textwidth,height=0.32\textwidth]{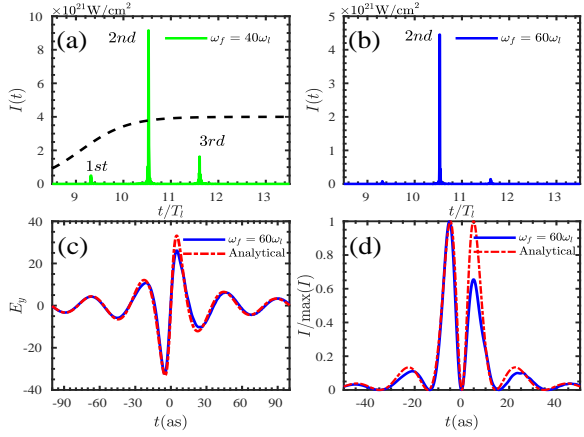}
 \caption{(color online) Attosecond pulses obtained by applying different filter frequency $\omega_f$: (a) $\omega_f=40\omega_l$, (b) $\omega_f=60\omega_l$. The upper frequency of the filter is $2000 \omega_l$. (c) Electric field $E_y$ and (d) normalized intensity of the $2$nd pulse compared with the analytical expressions. We label the pulse center at time $t=0$ and zoom in the time axis in unit of as. The laser, $a(t)=a_{0}(\tanh((t-T_s)/W)-\tanh((t-T_e)/W))/2$, radiates the plasma ($n_e=500n_c$, $L=\lambda_{l}/8$) with incident angle $\theta=\pi/4$, where $a_{0}=100$, $W=T_{l}=\lambda_{l}/c$, $T_s=6T_{l}$, $T_e=14T_{l}$, $\lambda_l=0.8\mu m$. The reflected laser profile (black dashed line) is shown in (a) with a.u.. The field detector is located at $3\lambda_l$ from the plasma surface.}\label{Fig.1.}
\end{figure}

\begin{figure}
  \includegraphics[width=0.48\textwidth,height=0.24\textwidth]{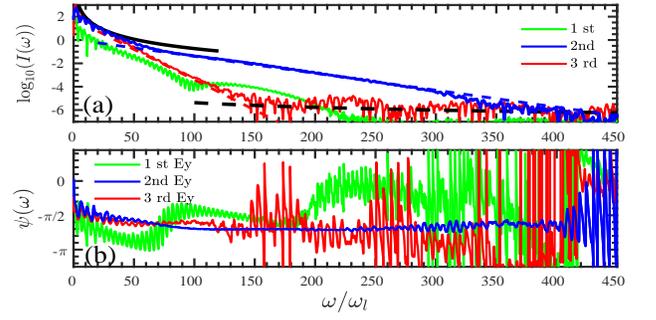}
 \caption{(color online) Intensity spectra (a) and spectral phase (b) for the pulses in Fig.\ref{Fig.1.}. The spectral fittings of the $2$nd pulse ($\textrm{log}_{10}(I)=0.1-0.0145\omega$, blue dashed line) and $3$nd pulse ($\textrm{log}_{10}(I)=1.9-0.052\omega$, red dashed line) are shown with the power-law spectral scalings ($I=10^{4.6}\omega^{-8/3}$, black solid, $I=10^{-2.7}\omega^{-4/3}$, black dashed) fitting the different regions in the spectra.}\label{Fig.2.}
\end{figure}

In Fig.\ref{Fig.1.}, we show the obtained AS pulses by filtering out low order harmonics ($\omega < \omega_f$) in the reflected wave from an ultra-dense plasma irradiated by a long duration laser pulse. One can clearly see the action of the plasma gate for pulse isolation as only three pulses are seen. Even though the laser is still on, no strong AS pulses can be emitted after the $3$rd pulse in Fig.\ref{Fig.1.} (a). Here the $1$st pulse arises due to the reflection of the laser ramp from the pre-plasma present at the target surface. The $2$nd pulse with intensity $I_2\approx 9.2\times10^{21} \textrm{W/cm}^{2}$ is emitted in the first period of the peak laser interacting with the bulk of the plasma target. During this period, a large part of the laser energy is first stored in the plasma electrostatic field due to the compression of the electrons into an ultra-dense nanometer layer. The stored energy is then released by backward accelerating the electron layer to emit an ultra-intense AS pulse. This energy conversion process is described by the RES model~\cite{Gonoskov:2011aa} and can emit a pulse with intensity stronger than the pulses reported so far~\cite{PRL245005CSE}. The $3$rd pulse generated in the next period is emitted with much weaker intensity than the $2$nd pulse. From this period, the HB effect becomes important, and the plasma gating starts to speed up the pulse spectrum decay and degrade the spectral phase coherence. One can see in Fig.\ref{Fig.2.} (a), the $2$nd pulse has a slower intensity decay than the $1$st and $3$rd pulses, implying higher efficiency for high frequency emission. In Fig.\ref{Fig.2.} (b), the high-frequency components in the $1$st and $3$rd pulses display larger phase fluctuation than that in the $2$nd pulse, which could further reduce the pulse intensity and extend the duration. In the following laser periods, the generated pulses would have much faster spectral decay and more fluctuated spectral phase. Thus, we can isolate the $2$nd pulse with a suitable frequency filter and enhance the isolation with a larger filtering frequency $\omega_f$, \emph{e.g.} $I_2/I_3=6.10$ for $\omega_f=40$ in (a), $I_2/I_3=31.13$ for $\omega_f=60$ in (b). In fact, this is in line with our previous results, the cut-off-frequency in (b) corresponds to Eq.(2) of~\cite{Tang:2017aa} for the parameters considered here.

To complement the simulation results, we also give the analytical calculations to depict the properties of the emitted pulse. We assume that the generated pulse is mainly radiated by the surface electron layer~\cite{Gonoskov:2011aa}, and at the emission instant $(t',x'(t'))$, the layer transverse momentum $p_{y}=\gamma\beta_y(t')$ passes through the zero node~\cite{PRL245005CSE}, and its backward velocity $\beta_x(t')\approx -1$ approaches the speed of light. The surface layer with nanometer thickness $\Delta x$~\cite{pop103301} can be approximated to $n_{el}\delta(x'-x'(t'))$ for the coherent emission if wavelength $\lambda_{\omega}\gg\Delta x$. Substituting these in $E_y^{r}(x,t)=-\int_{x}^{\infty} J_y(x^{'},t^{'})dx^{'}/2$, one can get $E^{r}_{y}(t)=-\hat{E^{i}_{y}}2A_m\omega_{d}t/(1+(\omega_{d}t)^2)$, where $\hat{E^{i}_{y}}$ is the sign of the laser electric field $E^{i}_{y}(x',t')$, $\omega_d=|\frac{dp_y}{dt'}|/(1+\beta_x)\approx2\gamma^{2}|\frac{dp_y}{dt'}|$, $A_m=n_{el}\gamma(1-\beta_x)/4\approx n_{el}\gamma/2$, $n_{el}(t')$ is areal density of the electron layer. The retardation relation $t'+x'(t')=t+x$ is used and the field detector at $x$ is far enough from the plasma surface. For convenience, we label the pulse center at $t=0$. On filtering out the low frequency components ($\omega<\omega_f$), we obtain an AS pulse:
\begin{equation}
 E^{r}_{y}(\omega_f,t)=\frac{-2A_m}{\sqrt{1+(\omega_dt)^2}}e^{-\frac{\omega_f}{\omega_d}}\cos(\omega_{f}t+\varphi(t))\hat{E^{i}_{y}}
\label{eq1}
\end{equation}
with the phase chirp $\varphi(t)$: $\cos(\varphi)=\omega_dt/\sqrt{1+(\omega_dt)^2}$, $\sin(\varphi)=-1/\sqrt{1+(\omega_dt)^2}$. This formula is applicable for oblique incidence by treating in a Lorentz boosted frame~\cite{Bourdier:1983aa} with $A_m \approx n_{el}\gamma\cos^{-2}(\theta)/2$. The pulse duration $T_d=2/\omega_d$ is extremely shortened with the relativistic backward velocity and larger transverse acceleration. The pulse is characterized by an exponential spectrum~\cite{Gonoskov:2011aa}, $I(\omega)=|E^{r}_{y}(\omega)|^2=A^{2}_{m}/\omega^{2}_{d}\exp{(-2\omega/\omega_d)}$. In Fig.\ref{Fig.2.} (a), we confirm this exponential spectrum with the linear-logarithm fitting, $\log_{10}(I(\omega))=\log_{10}(A_m^2/\omega_d^2)-\log_{10}(e^2)\omega/\omega_d$, and find that the exponential region is bounded by the $\gamma$-spike scaling $I(\omega)\propto \omega^{-8/3}$~\cite{Baeva:2006aa} in low-frequency region and CSE scaling $I(\omega) \propto \omega^{-4/3}$~\cite{Brugge:2010aa} in high-frequency region. The fitting slope of the 2nd pulse spectrum reveals the spectral decay $1/\omega_d=0.017$ and precisely gives the pulse duration $T_d=14.2 \textrm{as}$ which is much shorter than the duration of the $3$rd pulse ($1/\omega_d=0.060$, $T_d=50.8 \textrm{as}$). With Eq.(\ref{eq1}) and $A_m \approx 57$ obtained from the simulation, the $2$nd pulse is reproduced in Fig.\ref{Fig.1.}(c), (d) for filtering frequency $\omega_f = 60 \omega_l$. Another salient feature of the pulse is the constant spectral phase, $\psi_0(\omega)=-\pi/2$ (or $\pi/2$ if $E^{i}_{y}$ changes sign). This particular locked phase is the consequence of the transverse current changing its sign at the node where the transverse momentum becomes zero. This also results in a minimum at the center of the pulse, contrary to a synchrotron-like pulse~\cite{PRL245005CSE}. We stress that the locked phase does not depend on the CEP of the long laser pulse, but on the dynamics of the well-defined electron layer during the emission. In Fig.\ref{Fig.2.}(b) we quantitatively confirm the constant spectral phase in $E_y$, $\psi(\omega)\approx-\pi/2$ and in $B_z$, $\psi(\omega)\approx\pi/2$ as propagating in $-x$ direction (not shown). The phase mismatch in lower and higher frequency regions may come respectively from the interference with the emissions from ROM/CWE and CSE. We also note that in Fig.\ref{Fig.2.} the constant phase deviates slightly from $\psi_0(\omega)$. This deviation $\psi_{A_m}\sim A'_{m}/(A_m\omega_d)$, resulting from the variation of $A_m(t')$ during the pulse emission, causes the asymmetry of the emitted pulse in Fig.\ref{Fig.1.} (d), where $A'_{m}=\frac{dA_m(t')}{dt'}/(1+\beta_x(t'))$ is the temporal derivative of $A_m$. In the high frequency region, the phase fluctuation occurs. This phase fluctuation, $\psi_f(\omega) \sim 2\pi\Delta x/\lambda_{\omega}$, originates from the finite extension of the surface layer and may reflect the incoherence of the high frequency emission from different part of the layer if $\lambda_{\omega} \lesssim \Delta x$.  We verify this by calculating the layer thickness $\Delta x$ in the simulations and observing the threshold of the phase fluctuation $\omega^{th}_{f}\approx 2\pi c/\Delta x$, \emph{e.g.} $\Delta x\approx0.002\lambda_l$, $\omega^{th}_{f} \approx 400\omega_l$ for the $2$nd pulse in Fig.\ref{Fig.2.}(b).

\begin{figure}
  \includegraphics[width=0.48\textwidth,height=0.40\textwidth]{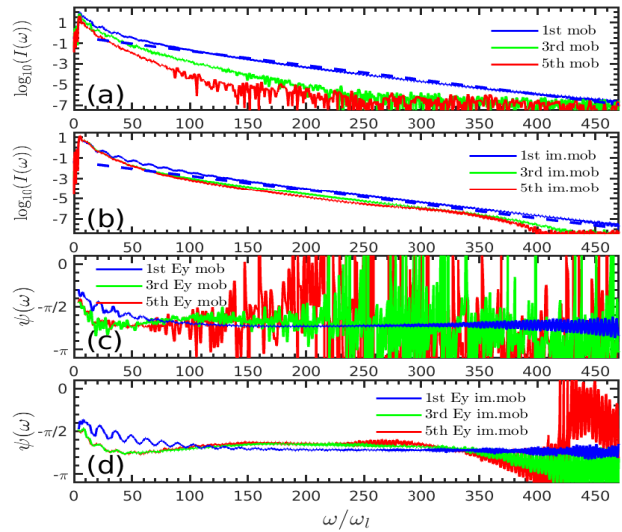}
  \caption{(color online) Intensity spectra (a) (b) and the corresponding spectral phase (c) (d). (a), (c) correspond to the case with mobile ions and (b), (d) immobile ions. The blue dashed lines are the spectral fitting for the $1$st pulses in both cases. The laser, $a_0=40$ with step-like temporal profile, radiates normally on the plasma ($n_e=200n_c$, $L=0$), and the laser duration is long enough to emit the $5$th pulse.}\label{Fig.3.}
\end{figure}

With these simulation results and analytical calculations, we can obtain a comprehensive understanding of the plasma gating. As discussed, the pulse amplitude $A_m$ and spectral decay $1/\omega_d$ depend sensitively on $\gamma$ which is determined by the backward acceleration of the electron layer due to the charge separation field. In HB evolution, the charge separation is mitigated because the Doppler effect decreases the laser pressure, and part of the energy in the electrostatic field is absorbed by mobile ions. The layer acceleration thus would be restricted by the HB effect, leading to a smaller amplitude and faster decay for the $3$rd pulse, for which the HB effect is more important. Simultaneously, the HB effect inevitably spreads the structure of the backward-moving electron layer and thus reduces the phase fluctuation threshold $\omega^{th}_{f}$, which shortens the coherent phase interval and further decreases the amplitude of the $3$rd pulse. Essentially, the HB effect isolates the $2$nd pulse by suppressing the coherent emission for the $3$rd pulse. To demonstrate the isolating effect of the HB evolution, we compare the pulse spectra and the spectral phase in the cases with mobile and immobile ions, where almost pure HB effect is in action. For the mobile-ion case, the decay of the pulse spectra in Fig.\ref{Fig.3.} (a) becomes faster and faster, while in Fig.\ref{Fig.3.} (b) for the immobile case, the pulse spectra sustain the same decay. In Fig.\ref{Fig.3.} (c), the more phase fluctuation is induced in the emitted pulses with HB effect than that in Fig.\ref{Fig.3.} (d) for immobile case. Another effect also contributing to the pulse isolation is plasma heating. It expands the layer and decreases the number of electrons for coherent emission. In Fig.\ref{Fig.3.} (b), the following pulses have the same spectral decay as the $1$st one but weaker intensity. This is because the pulse emission process is roughly repeatable since ions are fixed, but the layer compression becomes less in longer time interaction. In Fig.\ref{Fig.3.} (c), the threshold of the phase fluctuations is reduced by thermal expansion. With oblique incidence and long plasma gradient in Fig.\ref{Fig.1.}, the plasma heating is more considerable with the re-injection of Brunel electrons~\cite{Brunel} which could effectively disperse the layer structure and consequently make larger phase fluctuations in the pulse. For the $1$st pulse in Fig.\ref{Fig.1.}, the plasma heating is strong in the interaction of the laser ramp with pre-plasma. Hence, a well-defined electron layer can't be formed leading to weaker intensity in Fig.\ref{Fig.2.} (a) and large phase fluctuations in Fig.\ref{Fig.2.} (b). All of these effects, \emph{e.g.} HB effect, plasma heating, contributes to the plasma gating and isolates the $2$nd pulse in Fig.\ref{Fig.1.} with the most efficient high-frequency emission and the widest coherent phase interval.

\begin{figure}
  \includegraphics[width=0.48\textwidth,height=0.20\textwidth]{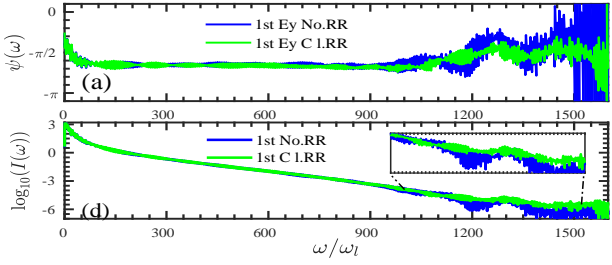}
  \caption{(color online) Spectral phase (a) and intensity spectrum (b) for the cases with and without RR force. Same parameters in Fig.\ref{Fig.1.} except $a_{0}=250$, $W=0T_{l}$, $n_e=1000n_c$.}\label{Fig.4.}
\end{figure}

\begin{figure}
  \includegraphics[width=0.48\textwidth,height=0.40\textwidth]{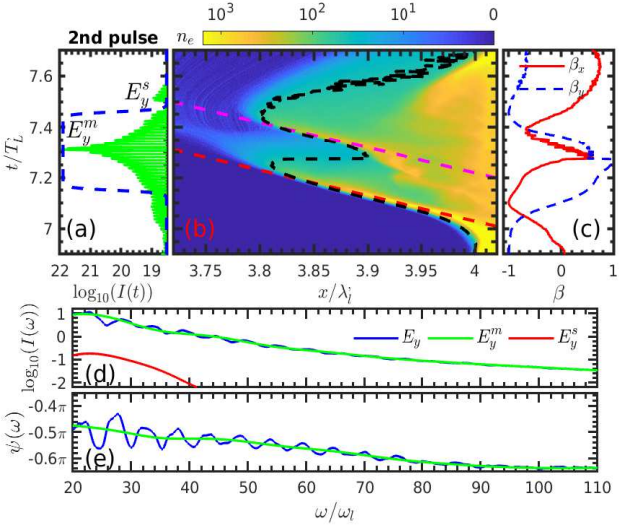}
  \caption{(color online) (a) Temporal shape of the $2$nd pulse in Fig.\ref{Fig.1.} (a) shifted along the retardation relation. (b) Evolution of the electron density $n_e$ at the plasma surface around the emission of the $2$nd pulse overlaid with the retardation paths of the pulse centers. The evolution of the electron surface ($n_e=a_0n'_c$) is also shown (black dashed line). (c) Velocity ($\beta_x$,$\beta_y$) of the electron surface. (d) Spectra of the main pulse (green line) and sub-pulse (red line). (e) Spectral phase of the main pulse (green line). The spectrum and spectral phase in Fig.\ref{Fig.2.} of the $2$nd pulse in the low frequency region are repeated in (d) and (e) respectively. $\lambda'_l=\lambda_l/\cos(\theta)$, $T'_l=T_l/\cos(\theta)$, $n'_c=n_c\cos^{2}(\theta)$ are defined in the Lorentz boosted frame~\cite{Bourdier:1983aa}}\label{Fig.5.}
\end{figure}

The phase fluctuation in high frequency region limits the interval of coherent emission. In order to extend the constant phase interval to higher frequency, an intenser laser pulse $(I\sim 10^{23}\,\textrm{W/cm}^{2})$ with shorter ramping front is used to compress the electron layer narrower. As shown in Fig.\ref{Fig.4.} (a), the threshold $\omega^{th}_f$ of the phase fluctuation is clearly improved with a stronger laser as the layer is further compressed, $\Delta x \sim (n_ea_0)^{-1/3}$~\cite{Gonoskov:2011aa}. Here the $1$st pulse becomes the main pulse because no laser ramp exists. In this ultra-relativistic case, we also find that in Fig.\ref{Fig.4.} (b), RR force reduces the phase fluctuation in high-frequency region. This may suggest the role of RR force in further compressing the electron layer. One can also see in Fig.\ref{Fig.4.} (b) the emitted pulse with RR force has the same spectral decay as the case without RR force but smaller spectral fluctuation in the region $\omega > 1000\omega_l$. This may be because the collective motion of the layer can hardly be impeded by RR force as the laser field can not penetrate into the ultra-dense electron layer. The spectral fluctuation comes from the superposition of the incoherent emissions from the layer and RR force smooths the fluctuation by further compressing the layer structure. With the same spectral decay and the smaller phase fluctuation, a stronger AS pulse might be synthesized. After a long time interaction, the electron layer will be extremely heated~\cite{NJP123005} and expanded. RR force would increase the energy absorption and decrease the total reflection~\cite{Tang:2017aa}.

In Fig.\ref{Fig.5.}, the spectral and phase oscillations in low-frequency region are clearly presented. The oscillation is the consequence of the interference between the double pulses in one emission process. Fig.\ref{Fig.5.} (a) shows that the $2$nd pulse in Fig.\ref{Fig.1.} (a) consists of a main pulse $E_y^m$ and a sub-pulse $E_y^s$, \emph{i.e.} $E_y(t)=E_y^m(t)+E_y^s(t-\Delta t)$, $E^m_y(t) \gg E^s_y(t)$, where $\Delta t \approx 0.18T_l$ is the time separation. In Fig.\ref{Fig.5.} (b), we plot the contour of the evolution of the electron density and the retardation paths of the two pulses. It confirms the assumptions that the (main) pulse is emitted by the surface nanometer electron layer at the node where $\beta_y \approx 0$ and $\beta_x \approx -1$ in Fig.\ref{Fig.5.} (c). The secondary electron bunch~\cite{Subpulse} formed behind the first electron layer radiates the sub-pulse with the much slower backward velocity, thus the efficiency of the high frequency emission is significantly lower than that in the main pulse in Fig.\ref{Fig.5.} (d). By artificially excluding the sub-pulse with the temporal window (blue dashed line in Fig.\ref{Fig.5.} (a)), one can see that in Fig.\ref{Fig.5.} (d) (e) the main pulse has no spectral and phase oscillations while the whole 2nd pulse which is a superimposition of the two pulses shows the oscillations. To describe the oscillations analytically, we calculate spectral and phase oscillations in the whole pulse from the double pulse interference, $|E_y(\omega)|e^{i\psi(\omega)}=|E_y^m(\omega)|e^{i\psi_m(\omega)} + |E_y^s(\omega)|e^{i\psi_s(\omega)}e^{i\omega\Delta t}$, and obtain
\begin{align}
 I(\omega)\approx & |E_y^m(\omega)|^2[1 + 2\frac{|E_y^s(\omega)|}{|E_y^m(\omega)|} \cos(\theta)]\nonumber\\
 \psi(\omega)\approx & \psi_m(\omega)+\frac{|E_y^s(\omega)|}{|E_y^m(\omega)|}\sin(\theta)
\label{eq2}
\end{align}
where $\theta(\omega)=\psi_s(\omega)-\psi_m(\omega)+\omega\Delta t$, $|E_y^{m,s}(\omega)|$ denotes the modulus of the different frequency components and $\psi_{m,s}$ is the pulse spectral phase. As we can see, the spectrum and phase oscillate qualitatively with the frequency $\omega_t=2\pi/\Delta t\approx5.56\omega_l$ which matches very well with the simulation results, and the oscillation amplitude attenuates for higher frequency because of the less efficient high frequency emission from the secondary electron bunch. 

In summary, we have shown the existence of a plasma gating to generate a ultra-intense single AS pulse with duration $T_d<20$as. Contrary to other schemes, the plasma gating is robust and works for general situations \emph{e.g.} plasma gradient, oblique incidence, long laser pulse driver. The harmonics constituting the generated pulse are phase-locked to $\psi(\omega)=\pm \pi/2+\psi_{A_m}$, and can be coherently extended to KeV region.  A promising application of the plasma gating is to coherently focus~\cite{PRL103903} the AS pulse boosting its intensity to $I_{\textrm{CHF}}=2(2R_0\Omega/\lambda_l)^{2}I_{l}\omega^{2}_d$, where $R_0$ and $\Omega$ are the radius and solid angle of the surface. With the same parameters ($R_0=4\lambda_l$, $\Omega=1$) as in~\cite{PRL103903}, Schwinger limit can be reached with an incident laser of intensity $I_{l}\sim3.6\times10^{23} \textrm{W/cm}^{2}$ which will be available in the ELI project\cite{eli:nn}.

We thank Prof. Christoph Keitel for his helpful advice and appreciation for this work.

\bibliography{atto_pulse,atto_pulse2,hhg,shocks}

\end{document}